\documentclass[journal,transmag]{IEEEtran}
\usepackage{cite}
\ifCLASSINFOpdf
\else
\usepackage[dvips]{graphicx}
\fi
\usepackage{amsmath,ragged2e,color,amsmath, amsthm, amssymb,amsfonts}
\usepackage{algorithmic}
\usepackage{array}
\usepackage{subfig}

\begin{document}

\title{Three-dimensional modulation instability of dust-ion-acoustic waves and rogue waves in warm nonthermal magnetized plasmas}

\author{\IEEEauthorblockN{A. Parvez\IEEEauthorrefmark{1,2}, A. Mannan\IEEEauthorrefmark{1,3}, M. N. Haque\IEEEauthorrefmark{4,*}, and A. A. Mamun\IEEEauthorrefmark{1}}
\IEEEauthorblockA{\IEEEauthorrefmark{1}Department of Physics, Jahangirnagar University, Savar, Dhaka-1342, Bangladesh}
\IEEEauthorblockA{\IEEEauthorrefmark{2}Department of Electrical \& Electronic Engineering, Uttara University, Uttara, Dhaka-1230, Bangladesh}
\IEEEauthorblockA{\IEEEauthorrefmark{3}Laboratori Nazionali di Frascati, INFN, Via Enrico Fermi 54, 00044, Frascati, RM, Italy}
\IEEEauthorblockA{\IEEEauthorrefmark{4}Department of Physics, South Dakota School of Mines and Technology, SD 57701, USA}
\IEEEauthorblockA{\IEEEauthorrefmark{*}Email: mdnurul.haque@mines.sdsmt.edu}
\thanks{}}

\IEEEtitleabstractindextext{
\begin{abstract}
A theoretical investigation has been made to study the modulation stability/instability of three-dimensional dust-ion-acoustic wave packets in the warm magnetized complex plasma system in the presence of nonthermal distributed electrons and positrons species. The set of equations describing our plasma system has been reduced to a (3+1)-dimensional nonlinear Schr\"odinger equation by using the reductive perturbation method that is valid for finite but small amplitude limits. It is observed that both nonlinear and dispersive coefficients of (3+1)-dimensional nonlinear Schr\"odinger equation are significantly modified by the external magnetic field and transverse velocity perturbation. The regions of stable and unstable for the modulated dust-ion-acoustic waves have been examined numerically. Moreover, the dependence of modulation instability and rogue waves on the relevant plasma parameters is discussed. The implications of our theoretical results in space and laboratories magnetized dusty plasma medium is briefly discussed.
\end{abstract}
}
\maketitle
\IEEEdisplaynontitleabstractindextext
\IEEEpeerreviewmaketitle
\section{Introduction}
\label{6sec:Introduction}
Nowadays, the study of magnetized plasma \cite{El-Awady2014,Guo2014,Abdelwahed2017,Sethi2018,Irfan2019,Haque2019,Haque2020,Haque2021,El-Taibany2022} is a very interesting topic among modern experimental and theoretical physicists. It is
because of the important applications in many fields, such as the semiconductor industry, rocket science, fusion reactor, and so on.
Also, a deep understanding of the magnetized plasma is essential to examine the nature of our universe, because
in our universe 99 present visible matters are plasma. When massive dust grains ($\mu m$ to sub-$\mu m$ in size) are present
in pure plasma then it is called dusty plasma (DP) and due to its natural behaviours, it is also called a very complex plasma system. Many researchers
studied the nonlinear dynamics in DP, such as modulation instability (MI) and natural properties of nonlinear waves (solitary wave,
shock wave, rouge wave, oscillatory structures, etc.). El-Taibany \textit{et al.} \cite{El-Taibany2022} have observed the Dust-acoustic (DA) solitary
waves in magnetized dusty plasmas, and numerically showed the relative dependency of DA solitary waves on various plasma parameters.
El-Labany \textit{et al.} \cite{El-Labany2021} have examined the DA waves, MI, and rouge waves (RWs) in multi-components DP medium with Maxwellian ions and double spectral electron distribution. They found that DA waves and RWs are very sensitive on two spectral indices, $r$ and $q$. The energy of RWs increases with $r$ but decreases with $q$.

The velocity distribution function (VDF) is one of the fundamental properties of plasma particles.
The nonlinear wave properties of magneto-plasma strongly depend on the plasma particle (electrons, positrons, ions, and so on) VDF.
Inside the plasma, due to the thermal energy, particles always collide with each
other, and exchange their momentum to balance the thermal equilibrium process.
Maxwellian VDF is a mainstream VDF to examine plasma particles. Nevertheless, many
observations have proved the presence of highly energetic particles (e.g., electrons and positrons)
in laboratories and space \cite{Vasyliunas1968,Feiter1973,Shrauner1979,Alinejad2004,Lashgarinezhad2017,Pakzad2009}. These excess energetic particles do not follow the
Maxwellian distribution function \cite{Vasyliunas1968,Feiter1973}.
Cairns \textit{et al.} \cite{Cairns1995} in 1995 proposed a VDF, known as Cairns nonthermal
distribution, to model the extra energetic particles that were noticed by the
Viking spacecraft \cite{Bostrom1992} and the Freja satellite \cite{Dovner1994}.
Under certain conditions, nonthermal distribution reduces to standard Maxwellian VDF.
Alinejad \textit{et al.} \cite{Alinejad2004} have studied the nonlinear ion-acoustic (IA) solitary waves (IASWs)
in a multi-components plasma with nonthermal electrons and observed that compressive and rarefactive IASWs are allowed to coexist
in the presence of nonthermal electrons. Also, positron density has a strong effect on the nonthermal
parameter. Lashgarinezhad \textit{et al.} \cite{Lashgarinezhad2017} have examined the consequence of nonthermal
positrons and electrons on IA cnoidal waves (IACWs) and found that IACWs amplitude increases with
nonthermal electrons, but decreases with nonthermal ions.
Pakzad \cite{Pakzad2009} has observed the influence of nonthermal electrons and positrons on the formation
of soliton.

The study of magnetosonic (MS) wave (MSW), which is a basic low-frequency mode in magnetized plasma \cite{El-Awady2014,Sethi2018},
is one of great interest among modern plasma physicists. The MSWs have importance in plasma
heating and accelerate the electrically charged particles \cite{Hazeltine2004,Ohsawa1985}, transport of energy in space and laboratory plasma \cite{El-Awady2014},
and so on. MSW is relatively similar to electromagnetic waves. Like electromagnetic wave MSW propagation direction is
perpendicular to the external magnetic force. The main purpose of this article is to examine the magnetosonic rouge waves (MSRWs) in
four components magnetized plasma. The MSRW is one of the most important nonlinear wave phenomena, which are first observed
in the Ocean and later in optics \cite{Solli2007,Shi2022}, magnetized plasma \cite{Haque2019,Haque2020,Haque2021,Chowdhury2017a}, hydrodynamics, and even in biology \cite{Turing1952}. It is a very
high energetic pulse, which develops in the unstable region to balance between stable and unstable regions. The characteristics
of MSRW are very complex and hard to mathematical modeling. Also, existing external forces like uniform
magnetic fields can alter the behaviors of MSRWs. So, it is very important to consider the external force to see the effects of it
on the formation and propagation properties of MSRW. Some authors \cite{El-Awady2014,Sethi2018,Haque2019,Haque2020,Haque2021,Chowdhury2017a,Haque2019a,Jahan2021} studied this type of wave in magnetized and unmagnetized
plasma media. Chowdhury \textit{et al.} \cite{Chowdhury2017a} have examined the dust-acoustic rogue (DAR) waves in unmagnetized space plasma and found
that propagation properties of DAR wave drastically changed with superthermal parameter and other plasma parameters.
El-Awady \textit{et al.} \cite{El-Awady2014} have studied the MS rogons in ion-electron plasma and found that dense plasma and strong
magnetic field are the cause of decreasing the nonlinearity of the medium, which significantly affects the rogons' amplitude.

In our present work, we study the modulation stability/instability of DIA waves in the presence of an external uniform magnetic field and modulation obliqueness in a three-dimensional warm nonthermal dusty plasma medium. It is observed that in the presence of transverse perturbation, the parametric regime characterizing the modulation instability regime is different from in the case of unmagnetized case. We also describe the properties of the first-order DIA rogue waves.

The manuscript is organized as follows. The 3D model equations are presented in Sec. \ref{6sec:Model and NLSE} and derivation of (3+1)-dimensional NLSE are reported in Sec. \ref{6sec:Derivation}.
The stability analysis of DIA waves and properties of DIA rogue waves are reported in Sec. \ref{6sec:MI and Rouge wave}. A summary is finally presented in Sec. \ref{6sec:Discussion}.

\section{Governing equations}
\label{6sec:Model and NLSE}
We consider a collisionless and four components three-dimensional complex magnetized plasma
medium whose constitutes are non-thermal electrons (mass $m_e$; charge $-e$) and positrons
(mass $m_p$; charge $+e$), static negatively charged dust grains (charge $q_d=-Z_de$), and
 warm ions (mass $m_i$; charge $q_i=+e$). At equilibrium the quasi-neutrality
condition is $N_{e0}+Z_dN_{d0}=N_{p0}+N_{i0}$, where $N_{s0}$ denotes the unperturbed
number densities of $s-$ species particles ($s=e,p,i,$ and $d$ for electrons, positrons,
ions, and dust, respectively); $e$ is the magnitude of the electron charge; $Z_d$ is the number
of electrons residing on a negative dust. We also assume an uniform external magnetic
field $\textbf{B}=B_0\hat{z}$ which acts along the $z$ axis, where $B_0$ is
the strength of the magnetic field and $\hat{z}$ the unit vector along the $z$-axis. Since the ion is assumed adiabatic,
the ion adiabatic index is $\gamma = (2+N)/N$, with $\gamma$ being the number
of degrees of freedom, which leads to $\gamma= 5/3$ by choosing $N=3$ for
three-dimensional cases. To describe the nonlinear dynamics of dust-ion-acoustic waves the dimensionless/normalized three-dimensional governing equations for magnetized dusty plasma medium are:
\begin{eqnarray}
&&\hspace*{-0.5cm} \partial_t n_i+\partial_x(n_iu_i)+\partial_y(n_iv_i)+\partial_z(n_iw_i)=0,
\label{6eq:1}\\
&&\hspace*{-0.5cm} \partial_tu_i+u_i\partial_xu_i+v_i\partial_yu_i+w_i\partial_zu_i=-\partial_x\Psi
\nonumber\\
&&\hspace*{4.0cm} -\chi_1\partial_xn_i^{2/3}-w_{ci}v_i,
\label{6eq:2}\\
&&\hspace*{-0.5cm} \partial_tv_i+u_i\partial_xv_i+v_i\partial_yv_i+w_i\partial_zv_i=-\partial_y\Psi
\nonumber\\
&&\hspace*{4.0cm} -\chi_1\partial_yn_i^{2/3}+w_{ci}u_i,
\label{6eq:3}\\
&&\hspace*{-0.5cm} \partial_tw_i+u_i\partial_xw_i+v_i\partial_yw_i+w_i\partial_zw_i=-\partial_z\Psi
\nonumber\\
&&\hspace*{4.5cm} -\chi_1\partial_zn_i^{2/3},
\label{6eq:4}\\
&&\hspace*{-0.5cm} \partial_{xx}\Psi+\partial_{yy}\Psi+\partial_{zz}\Psi=\chi_2n_e-\chi_3n_p-n_i
\nonumber\\
&&\hspace*{4.0cm} +\chi_3-\chi_2+1,
\label{6eq:5}
\end{eqnarray}
where $u_i$, $v_i$, and $w_i$ are the velocity components of the warm ion in the $x$, $y$, and $z$
direction, respectively, and normalized by the ion-acoustic speed $C_i=(k_BT_e/m_i)^{1/2}$ with $k_B$ being the Boltzmann constant and $T_e$ being the electron temperature; $n_s$ is number density of $s$-species normalized by its equilibrium density $N_{s0}$;
$t$ is the time variable normalized by the ion plasma frequency $\omega_{pi}^{-1}=(m_i/4\pi e^2N_{i0})^{1/2}$;
the space coordinates ($x$, $y$, $z$) are normalized by the Debye screening radius $\lambda_{Di}=(k_BT_e/4\pi e^2N_{i0})^{1/2}$;
$\Psi$ is electrostatic wave potential normalized by $k_BT_e/e$ and $\omega_{ci}=eB_0/m_i$ is
the ion cyclotron frequency normalized by $\omega_{pi}$; $\chi_1=5T_i/2T_e$ with $T_i$ is the ion temperature, $\chi_2=N_{e0}/N_{i0}$, and $\chi_3=N_{p0}/N_{i0}$.

The dimensionless non-thermal distributed electrons and positrons number densities can be written as \cite{Cairns1995,Alinejad2004}:
\begin{eqnarray}
&&\hspace*{0.0cm} n_e=\Big[1-\beta\Psi+\beta\Psi^2\Big]~\exp (\Psi),
\label{6eq:6}\\
&&\hspace*{0.0cm} n_p=\Big[1+\beta\chi_4\Psi+\beta\chi_4^2\Psi^2\Big]~\exp (-\chi_4\Psi),
\label{6eq:7}
\end{eqnarray}
where $\chi_4=T_e/T_p$ with $T_p$ is the temperature of positron, $\beta = 4\alpha/(1+3\alpha)$ and $\alpha$ defines the population of fast (energetic) particles in our plasma system.
For simplicity, we now substitute Eqs. \eqref{6eq:6} and \eqref{6eq:7} into Eq. \eqref{6eq:5}, and expand the
resulting equation up to third order as
\begin{equation}
\partial_{xx}\Psi+\partial_{yy}\Psi+\partial_{zz}\Psi+n_i-1=\chi_5\Psi+\chi_6\Psi^2+\chi_7\Psi^3+\cdot\cdot,
\label{6eq:8}
\end{equation}
where
\begin{eqnarray}
&&\hspace*{0.0cm} \chi_5=(1-\beta)(\chi_2+\chi_3\chi_4),
\nonumber\\
&&\hspace*{0.0cm} \chi_6=\frac{1}{2}(\chi_2-\chi_3\chi_4^2),
\nonumber\\
&&\hspace*{0.0cm} \chi_7=\frac{1}{6}(3\beta+1)(\chi_2+\chi_3\chi_4^3).
\nonumber\
\end{eqnarray}
\section{3+1 dimensional NLS equation}
\label{6sec:Derivation}
\begin{figure}[t!]
\centering
\includegraphics[width=80mm, height=70mm]{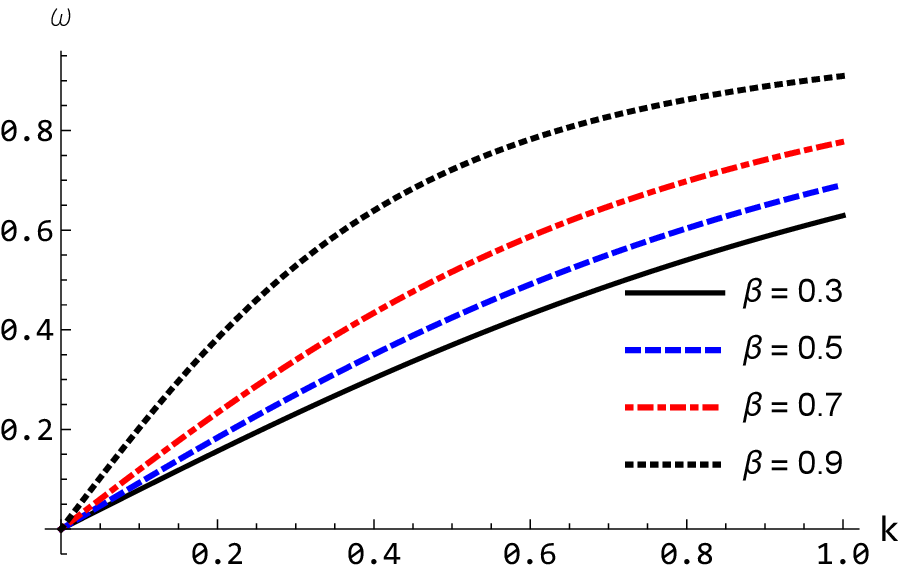}

\large{(a)}
 \vspace{0.5cm}

\includegraphics[width=80mm, height=70mm]{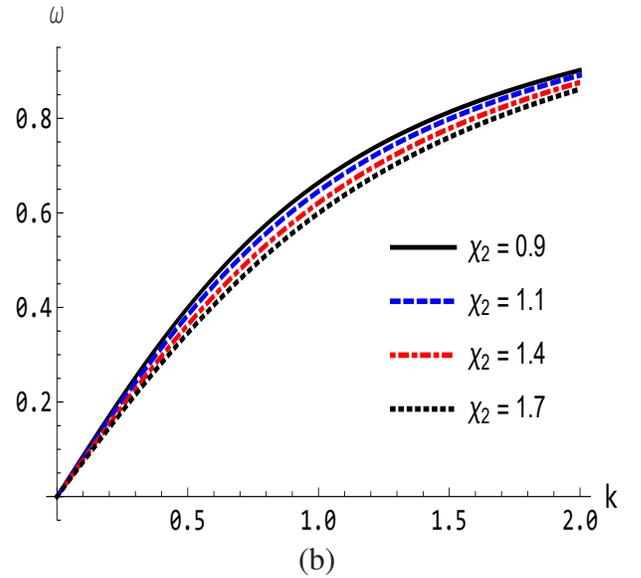}

\large{(b)}

\caption{The change of angular frequency $\omega$ with the carrier wave number $k$ for different values of : (a) $\beta=0.3, 0.5, 0.7$, and
$0.9$ when $\chi_2=1.3$; (b) $\chi_2=0.9, 1.1, 1.4$, and $1.7$ when $\beta=0.3$. The other plasma parameters are $\chi_1=0.025$, $\chi_3=0.8$, and $\chi_4=1.3$.}
 \label{6Fig:F1}
\end{figure}

To study the modulatonal instability and rogue waves, we here derive a $(3+1)$-dimensional NLSE by using the reductive perturbation method  \cite{Taniuti1969,Asano1969}. Therefore, we use the stretched independent variables as $\xi=\varepsilon x$, $\eta=\varepsilon y$, $\zeta=\varepsilon(z-v_gt)$, and $\tau=\varepsilon^2t$; where $v_g$ is the group velocity of dust-ion acoustic waves (DIAWs) and $\varepsilon~(0<\varepsilon<1)$ is a small parameter. Furthermore, the dependent variables $n_i$, $u_i$, $v_i$, $w_i$, and $\Psi$ are expanded as:
\begin{equation}\label{6eq:9}
\left.
\begin{aligned}
 & n_i=1+\sum\limits_{m=1}^\infty\varepsilon^{m}\sum\limits_{l=-\infty}^\infty n^{(m)}_{il}(\xi,\eta,\zeta,\tau)e^{i(kz-\omega t) l},\\
 & u_i=\sum\limits_{m=1}^\infty\varepsilon^{m+1}\sum\limits_{l=-\infty}^\infty u^{(m)}_{il}(\xi,\eta,\zeta,\tau)e^{i(kz-\omega t) l},\\
 & v_i=\sum\limits_{m=1}^\infty\varepsilon^{m+1}\sum\limits_{l=-\infty}^\infty v^{(m)}_{il}(\xi,\eta,\zeta,\tau)e^{i(kz-\omega t) l},\\
 & w_i=\sum\limits_{m=1}^\infty\varepsilon^{m}\sum\limits_{l=-\infty}^\infty w^{(m)}_{il}(\xi,\eta,\zeta,\tau)e^{i(kz-\omega t) l},\\
 & \Psi=\sum\limits_{m=1}^\infty\varepsilon^{m}\sum\limits_{l=-\infty}^\infty \Psi^{(m)}_l(\xi,\eta,\zeta,\tau)e^{i(kz-\omega t) l},
\end{aligned}
\right\}
\end{equation}
where the real variables $\omega$ ($k$) represents the ion angular frequency (carrier wave number). Since $n_i$, $u_i$, $v_i$, $w_i$
and $\Psi$ are real, the co-efficients in Eq. \eqref{6eq:9} satisfy the condition $A^{(m)}_{-l}=A_l^{(m)^\ast}$,
where $A=(\Psi,n_i, u_i, v_i, w_i)$ and the asterisk indicates the complex conjugate. Now expressing the equations \eqref{6eq:1}-\eqref{6eq:4} and \eqref{6eq:8} in terms of new variables $\xi,~\eta,~\zeta,$ and $\tau$, and inserting the equation \eqref{6eq:9} into the resulting equations, we can develop a different set of equations in various powers of $\varepsilon$. Therefore, we
obtain the first order quantities for $m=1$ and $l=1$ in term of first order electrostatic potential $\Psi^{(1)}_1$ in the matrix form as:
\begin{eqnarray}
&&\hspace*{-1.3cm} \Big[n^{(1)}_{i1},~w^{(1)}_{i1}\Big]^T=\Bigg[\frac{k^2}{\Pi^2},~\frac{k\omega}{\Pi^2}\Bigg]^T\Psi^{(1)}_1,
\label{6eq:10}
\end{eqnarray}
where $\Pi^2=\omega^2-k^2\chi_8$, $\chi_8=2\chi_1/3$, and $T$ denotes the transpose of the matrix. The relation between angular frequency ($\omega$) and carrier wave number ($k$) of DIAWs is obtained as:
\begin{eqnarray}
&&\hspace*{-1.3cm} \frac{\omega^2}{k^2}=\frac{1+k^2\chi_8+\chi_5\chi_8}{k^2+\chi_5}.
\label{6eq:11}
\end{eqnarray}
It is easily sees that the angular frequency depends on the nonthermal parameter $\beta$, carrier wave number (k), $s$-species temperature and number density, and the other plasma parameters. The variation of $\omega$ with $k$ for different values of $\beta$ and $\chi_2$ are displayed in Fig. \ref{6Fig:F1}. It is seen from Fig. \ref{6Fig:F1}(a) that the angular frequency of DIA waves increases with the nonthermal parameters $\beta$ and reaches a constant values with increasing the values of $k$. It is also observed that the angular frequency of DIA waves decreases with electron to ion number density ratio $\chi_2$ and increases with $k$, but for higher values of $k$ rate of changes of the angular frequency of DIA waves is very low (as shown in Fig. \ref{6Fig:F1}(b)).

\begin{figure}[t!]
\centering
\includegraphics[width=80mm, height=70mm]{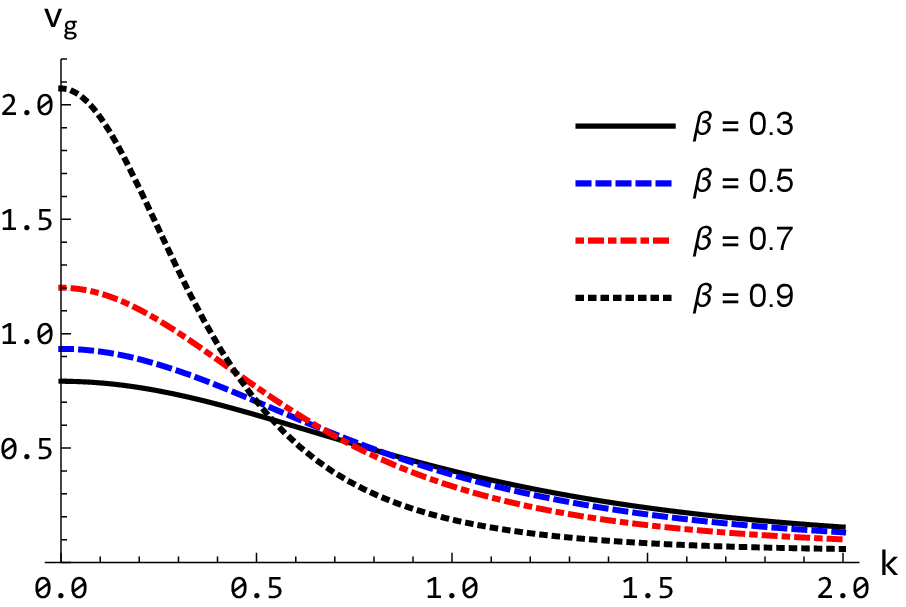}

\large{(a)}
 \vspace{0.5cm}

\includegraphics[width=80mm, height=70mm]{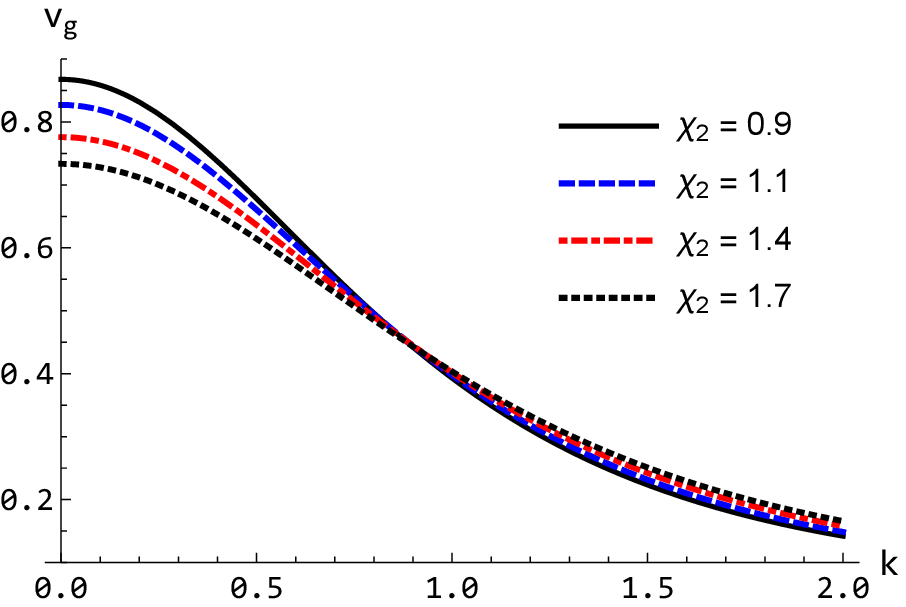}

\large{(b)}

\caption{The variation of group velocity $v_g$ with carrier wave number $k$ for different values of: (a) $\beta=0.3, 0.5, 0.7$, and
$0.9$ when $\chi_2=1.3$; (b) $\chi_2=0.9, 1.1, 1.4$, and $1.7$ when $\beta=0.3$. The other plasma parameters are $\chi_1=0.025$, $\chi_3=0.8$, and $\chi_4=1.3$.}
 \label{6Fig:F2}
\end{figure}

The equations for second-order $(m=2)$ and second harmonic components $(l=1)$ gives the following relations:
\begin{eqnarray}
&&\hspace*{-.8cm} i\omega w_{i1}^{(2)}=ik\Psi_1^{(2)}+i k\chi_8n_{i1}^{(2)}+\frac{(\omega^2-v_gk\omega)}{\Pi^2}\partial_{\zeta}\Psi_1^{(1)},
\nonumber\\
&&\hspace*{-.8cm} -i\omega n_{i1}^{(2)}=-ik w_{i1}^{(2)}+\frac{k(v_gk-\omega)}{\Pi^2}\partial_{\zeta}\Psi_1^{(1)},
\nonumber\\
&&\hspace*{-.8cm} n_{i1}^{(2)}+2ik\partial_{\zeta}\Psi_1^{(1)}=(k^2+\chi_5)\Psi_1^{(2)},
\nonumber\
\end{eqnarray}
and the above three equations give a compatibility condition as:
\begin{equation}\label{6eq:12}
v_g\left(=\partial_k\omega\right)=\frac{\omega^2-\Pi^4}{k\omega}\,.
\end{equation}
The corrections of the first-order quantities are
\begin{eqnarray}
&&\hspace*{-1.3cm} i\omega u_{i1}^{(1)}=-\omega_{ci}v_{i1}^{(1)}+\frac{\omega^2}{\Pi^2}\partial_{\xi}\Psi_1^{(1)},
\nonumber\\
&&\hspace*{-1.3cm} i\omega v_{i1}^{(1)}=\omega_{ci}u_{i1}^{(1)}+\frac{\omega^2}{\Pi^2}\partial_{\eta}\Psi_1^{(1)}.
\nonumber\
\end{eqnarray}
Equation \eqref{6eq:12} represents the group velocity of the DIAWs and no wave can propagate when the group velocity is zero. From Eq. \eqref{6eq:12} it is clear that the group velocity ($v_g$) of the DIAWs depends on the ion to electron temperature ratio, wave number $k$, angular frequency, and other plasma parameters. Figure \ref{6Fig:F2} reveals how the angular frequency of DIAWs changes with wave propagation constant $k$. It is observed that the group velocity decreases with the increasing the values of wave propagation vector and reaches a constant values as $k$ goes higher values. It is also observed that the value of $v_g$ increases (decreases) with nonthermal parameter $\beta$ ($\chi_2$), but at certain values of $k$ the rate of changes of $v_g$ with $k$ is very high for the higher (lower) values of $\beta$  ($\chi_2$).

The reduced equations for the second-order harmonic modes ($m=2$, $l=2$) can be written as:
\begin{eqnarray}
\begin{pmatrix}
              n^{(2)}_{i2} \\
              \\
              w^{(2)}_{i2}\\
              \\
              \Psi^{(2)}_2
\end{pmatrix}=
\begin{pmatrix}
              \chi_{11}^{(22)} \\
              \\
              \chi_{12}^{(22)}\\
              \\
              \chi_{13}^{(22)}
\end{pmatrix}|\Psi^{(1)}_1|^2,
\label{6eq:13}
\end{eqnarray}
where
\begin{eqnarray}
&&\hspace*{-0.3cm} \chi_{11}^{(22)}=\frac{3\omega^2k^4-k^6\chi_9+2\Pi^4k^2\chi_{13}}{2\Pi^6},
\nonumber\\
&&\hspace*{-0.3cm} \chi_{12}^{(22)}=\frac{2\Pi^4\omega^2k\chi_{13}+k^3\omega^4+2k^5\omega^2\chi_8-k^7\chi_8\chi_9}{2\Pi^6\omega},
\nonumber\\
&&\hspace*{-0.3cm} \chi_{13}^{(22)}=\frac{3\omega^2k^4-k^6\chi_9-2\Pi^6\chi_6}{6\Pi^6k^2},
\nonumber\
\end{eqnarray}
where, $\chi_9=\chi_8/3$.
The second-order quantities in the zeroth harmonic are found for $m=3$, $l=0$ and $m=2$, $l=0$
in term of $\Psi^{(1)}_1$ as:
\begin{eqnarray}
\begin{pmatrix}
              n^{(2)}_{i0} \\
              \\
              w^{(2)}_{i0}\\

              \\
              \Psi^{(2)}_0
\end{pmatrix}=
\begin{pmatrix}
              \chi_{14}^{(20)} \\
              \\
              \chi_{15}^{(20)}\\
              \\
              \chi_{16}^{(20)}
\end{pmatrix}|\Psi^{(1)}_1|^2,
\label{6eq:14}
\end{eqnarray}
where
\begin{eqnarray}
&&\hspace*{-0.4cm} \chi_{14}^{(20)}=\frac{\Pi^4\chi_{16}+2v_gk^3\omega+k^2\omega^2-k^4\chi_9}{\Pi^4(v_g^2-\chi_8)}
\nonumber\\
&&\hspace*{-0.4cm} \chi_{15}^{(20)}=\frac{\Pi^4\chi_{16}+\Pi^4\chi_8\chi_{14}+k^2\omega^2-k^4\chi_9}{v_g\Pi^4},
\nonumber\\
&&\hspace*{-0.4cm} \chi_{16}^{(20)}=\frac{2\chi_{6}(v_g^2-\chi_8)\Pi^4-2v_gk^3\omega-k^2\omega^2+k^4\chi_9}{\Pi^4(1-(v_g^2-\chi_8)\chi_5)}.
\nonumber\
\end{eqnarray}

Finally, after tedious but simple algebraic manipulations we obtain a $(3+1)$-dimensional NLSE for $m=3$ with $l=1$ in the following form
\begin{equation}\label{6eq:15}
 i\partial_{\tau}\Phi+P\partial^2_{\zeta\zeta}\Phi+Q|\Phi|^2\Phi-M\left(\partial^2_{\xi\xi}\Phi+\partial^2_{\eta\eta}\Phi\right)=0\,.
\end{equation}
Here, we assume $\Phi=\Psi^{(1)}_1$ for simplicity and the coefficients of $\Psi_1^{(3)}$ and $\partial_{\zeta}\Psi^{(2)}_1$ vanish because of expression of the group velocity and dispersion relation. $P$ and $M$ both are the dispersion coefficients and $Q$ is the nonlinear coefficient, respectively. The dispersive term $M$ contains the combined effect of ion angular frequency and ion cyclotron frequency. The dispersive and nonlinear terms are as follows:
\begin{eqnarray}
&&\hspace*{-0.3cm} P\Big(=\frac{1}{2}\partial_k v_g\Big)=-\frac{\Pi^2}{2k^2\omega^3}\Big(k^6\chi_8^3-5k^2\chi_8
\nonumber\\
&&\hspace*{0.9cm} \times\big(1+k^2\chi_8\big)\omega^2+\big(3+7k^2\chi_8\big)\omega^4-3\omega^6\Big)
\nonumber\\
&&\hspace*{-0.4cm} Q=\frac{3\Pi^4\chi_7}{2k^2\omega}-k\Big(\chi_{12}^{(22)}+\chi_{15}^{(20)}\Big)+\frac{\Pi^4\chi_6\Big(\chi_{13}^{(22)}+
\chi_{16}^{(20)}\Big)}{k^2\omega}
\nonumber\\
&&\hspace*{1.0cm} +\Big(\frac{k^2\chi_9}{2\omega}-\frac{\omega}{2}\Big)\Big(\chi_{11}^{(22)}+\chi_{14}^{(20)}\Big)-\frac{2k^6\chi_1}{27\Pi^4\omega},
\nonumber\\
&&\hspace*{0.3cm} M=\frac{\omega^4-\Pi^4\big(\omega^2-\omega_{ci}^2\big)}{2k^2\omega\big(\omega_{ci}^2-\omega^2\big)}.
\nonumber\
\end{eqnarray}
\section{Modulation Instability of DIAWs and Rogue Waves}
\label{6sec:MI and Rouge wave}
\begin{figure}[t!]
\centering
\includegraphics[width=75mm]{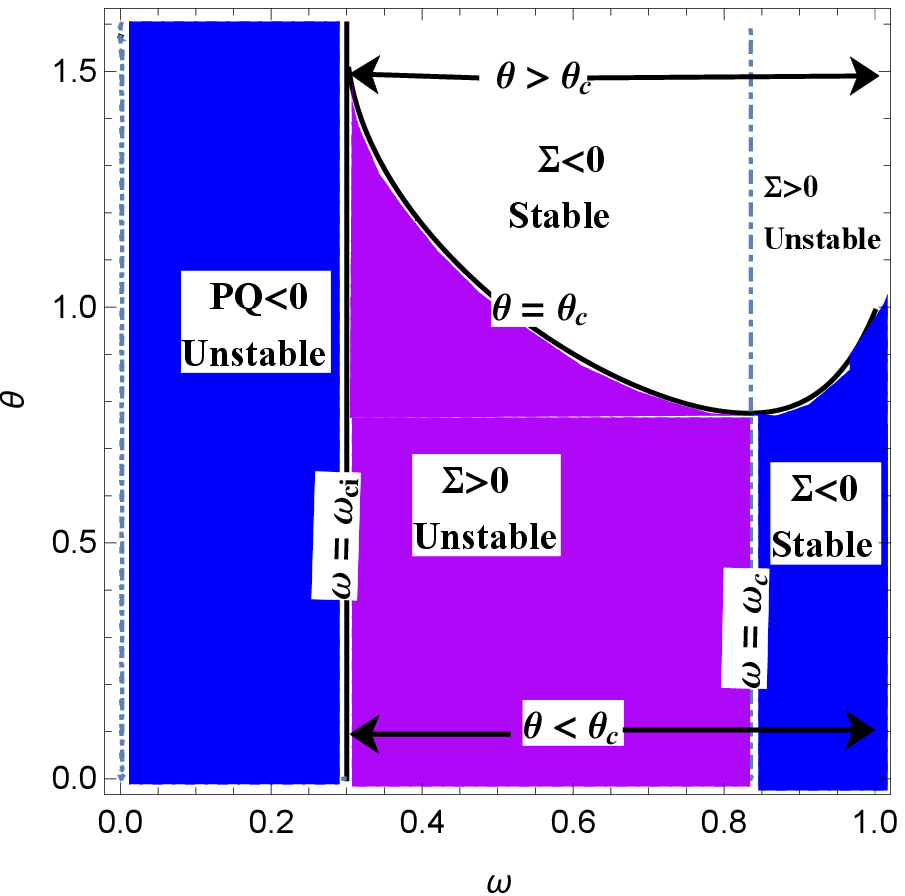}

\large{(a)}
 \vspace{0.5cm}

\includegraphics[width=75mm]{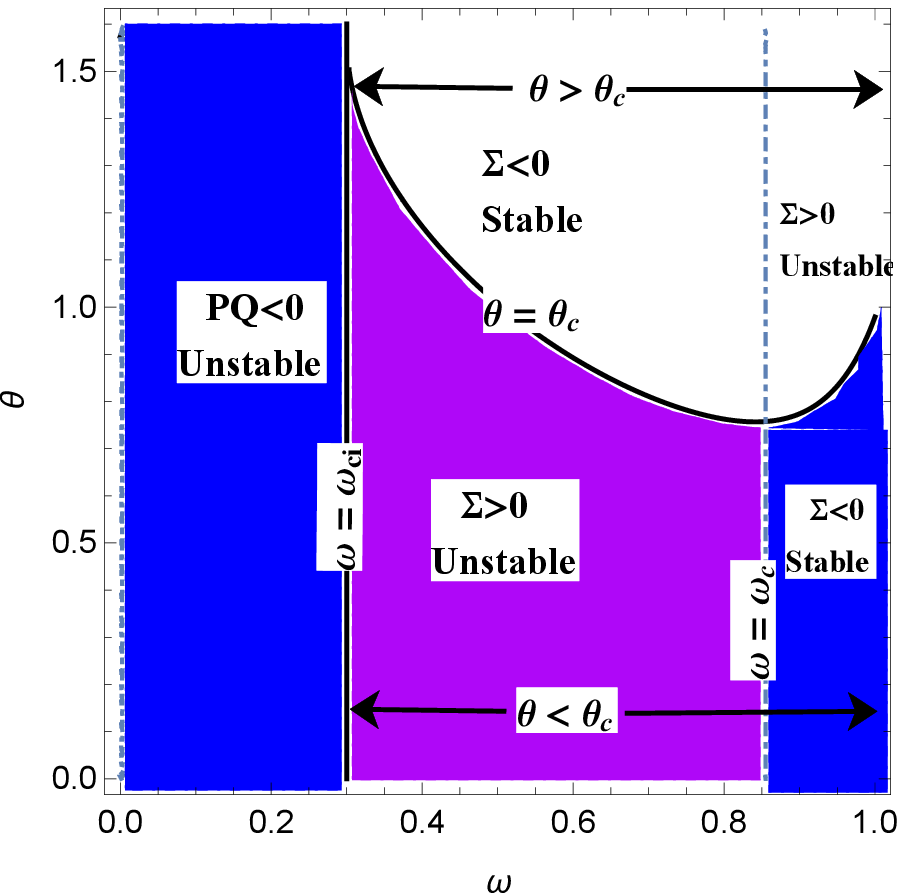}

\large{(b)}

\caption{The stable and unstable frequency regions of DIAWs in the $(\omega,\theta)$- plane for different values of $\beta$: (a) $\beta=0.3$ and (b) $\beta=0.6$. The other plasma parameters are $\chi_1=0.025$, $\chi_2=1.3$, $\chi_3=0.8$, $\chi_4=1.3$, and $\omega_{ci}=0.3$.}
\label{6Fig:F4}
\end{figure}
Now, we will discuss the MI of DIAWs and the properties of rogue waves in a magneto-complex plasma medium. To study the stability and instability of DIAWs, we will derive a dispersion relation from 3D NLSE Eq. \eqref{6eq:15}, and also find a rogue wave solution in the instability domain. It is well known that the MI of DIAW in an unmagnetized $1$D case (i.e. in the absence of transverse effect, $M=0$) is determined by the product $PQ$. But, the product $PQ$ in the 3D magneto-plasma system is not sufficient to describe the stability/instability frequency regimes. Further, the MI describing the frequency regimes of DIAWs in a 3D magneto-plasma system depends on longitudinal/transverse dispersion and nonlinear terms.

To study the modulational instability of DIA waves, we consider a plane wave solution of \eqref{6eq:15} with small modulation $\delta \Phi$ as
\begin{equation}\label{plane}
  \Phi = \left(\Phi_0 + \delta \Phi\right)e^{iQ|\Phi_0|^2\tau}\,,
\end{equation}
where $\Phi_0$ denotes the real constant that represents the amplitude of carrier wave, $|\delta\Phi|$ is much smaller than $\Phi_0$, and $-Q|\Phi_0|^2$ is the nonlinear frequency shift. Now, substituting \eqref{plane} into \eqref{6eq:15} we obtain the nonlinear dispersion relation for the amplitude modulation of DIA waves as \cite{Guo2014}
\begin{equation}\label{6eq:16}
 \Omega^2=K^4\left(\frac{\gamma_2/P}{1+\Pi_{\theta}^2}\right)^2\times\left(1-\frac{2|\Phi_0|^2
\big(1+\Pi_\theta^2\big)\gamma_1}{K^2\gamma_2}\right)\,,
\end{equation}
where $\gamma_1=PQ$, $\gamma_2= P^2(\Pi_\theta^2-M/P)$, $\Omega$ is the modulated wave frequency, and $K$ \Big($\equiv \sqrt{K_\xi^2+K_\eta^2+K_\zeta^2}$\Big) is the modulated wave number. Note that $K_\xi$, $K_\eta$, $K_\zeta$ are components of modulated wave number
$K$ along the space coordinates $\xi$, $\eta$, $\zeta$, respectively and the term $\Pi_\theta \Big(= K_\xi/\sqrt{K_\eta^2+K_\zeta^2}\Big)$
is connected with the modulational obliqueness $\theta(=\arctan(\Pi_\theta))$ which represents the angle between wave
vector $\mathbf{K}$ and the resultant $K_\zeta\zeta$ and $K_\eta\eta$. It is easily seen that the stability/instability of DIA wave packets depends on the factor of $2|\Phi_0|^2\left(1+\Pi_\theta^2\right)\Sigma$, where $\Sigma=\gamma_1/\gamma_2$. Therefore, \eqref{6eq:16} indicates that a critical wave number $\left(K_c = 2|\Phi_0|^2\big(1+\Pi_\theta^2\big)\Sigma\right)$ exists. Thus, modulational instability in three-dimensional evolution occurs for condition $K_c^2>K^2$ if one of the following conditions is satisfied: $\gamma_1<0$ and $\gamma_2<0$ or $\gamma_1>0$ and $\gamma_2>0$. This condition also allows that $\theta$ has a critical value
through $\theta_c\equiv \arctan\big(\sqrt{M/P}\big)$. It is concluded that the modulation unstable region ($\Sigma > 0$) exists when $\gamma_1$ and $\gamma_2$ are the same sign and the modulated stable region ($\Sigma<0$) exists with the opposite sign of $\gamma_1$ and $\gamma_2$. It is also important to note that for the condition $K_c^2 > K^2$ the decay/growth rate can be obtained from \eqref{6eq:16} as
$\Xi=Im~\Omega=(K^2\gamma_2/P+P\Pi_{\theta}^2)\times\sqrt{K^{-2}K_c^2-1}$. The maximum decay/growth rate is obtained as $\Xi_{max}=Q~|\phi_0|^2$ if the condition $PK_\xi^2-M(K_\eta^2+K_\zeta^2)=Q~|\Phi_0|^2$ is satisfied.

Figure \ref{6Fig:F4} shows the numerical analysis of MI. There are stability and instability frequency regimes, which are separated by the lines of $\omega = \omega_c$, $\omega= \omega_{ci}$, and $\theta=\theta_c$ in $\omega-\theta$ plane as illustrated in Fig \ref{6Fig:F4}. We see that the transverse dispersive term $M$, which depends on the value of $\omega_{ci}$, can be either negative or positive. The value of $M$ is undefined at $\omega=\omega_{ci}$. The dispersive term $M$ is positive (negative) at $\omega<\omega_{ci}$ ($\omega>\omega_{ci}$). On the other hand, the longitudinal dispersion term $P$ is always negative. The nonlinear term $Q$ is also be negative or positive. The nonlinear coefficient $Q$ vanishes at $\omega=\omega_c$. The value of $Q$ is positive (negative) at $\omega<\omega_c$ ($\omega>\omega_c$). For $0<\omega<\omega_c$, the product of $PQ$ is negative and the MI does not depend on the obliqueness $\theta$. On the other hand, for $\omega>\omega_{ci}$ the MI is strongly dependent of the obliqueness $\theta$. Therefore, the lines $\theta=\theta_c$ and $\omega=\omega_c$ divide $\omega-\theta$ the plane into four stable/unstable frequency regimes. It is worth mentioning that the value of the critical wave frequency $\omega_c$ increases with nonthermal parameter $\beta$. On the other hand, the critical wave frequency $\omega_c$ decreases with the electron to ion number density ratio $\chi_2$.
\begin{figure}[t!]
\centering
\includegraphics[width=75mm]{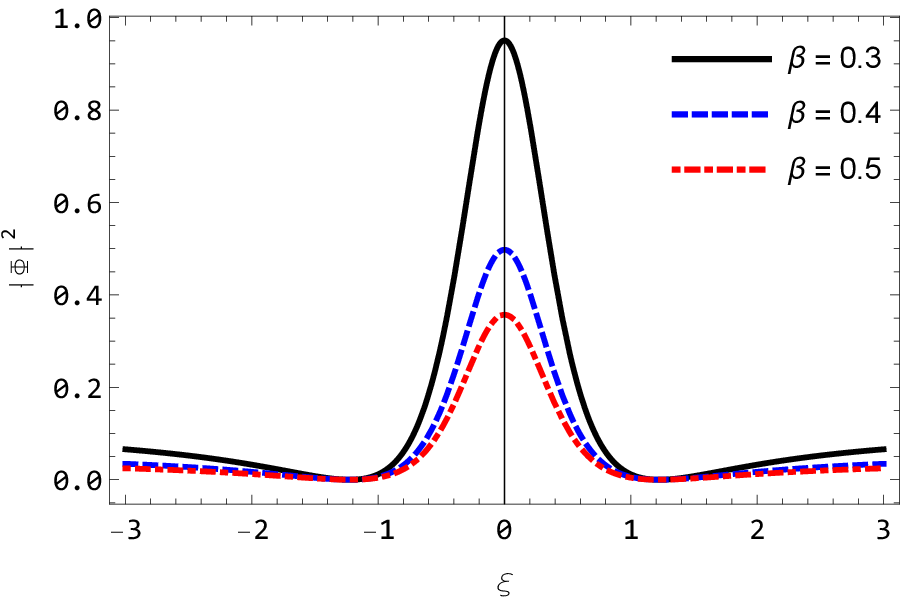}

\large{(a)}
 \vspace{0.5cm}

\includegraphics[width=75mm]{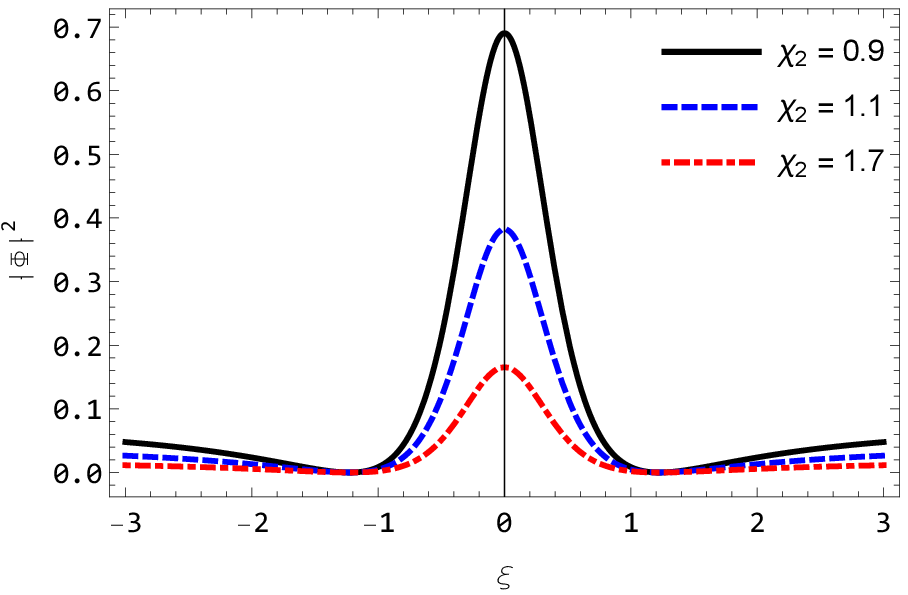}

\large{(b)}
 \vspace{0.5cm}

\includegraphics[width=75mm]{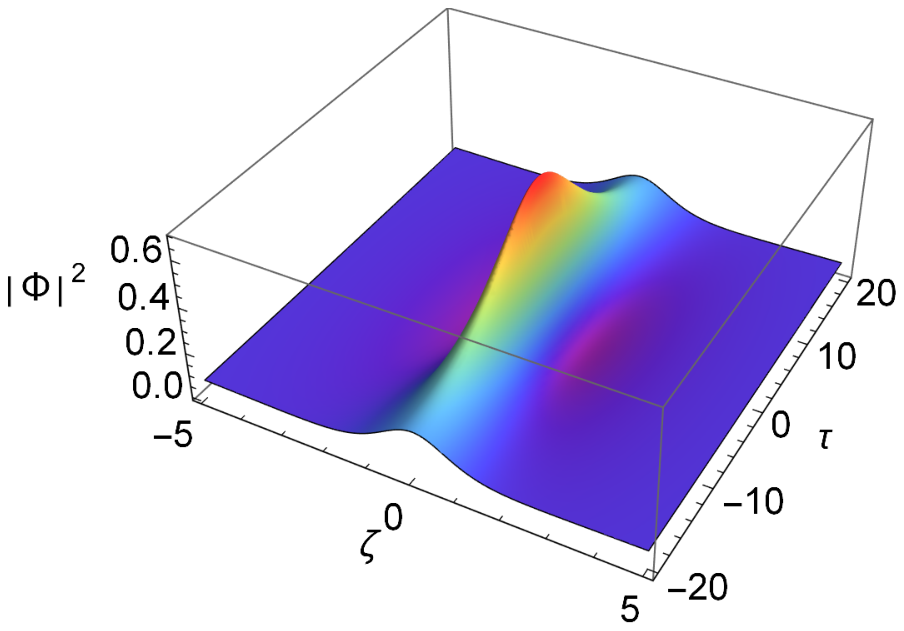}

\large{(c)}

\caption{The variation of amplitude and width of first-order rogue waves for different values of (a) $\beta=0.3, 0.4, \text{and}~ 0.5$ when $\chi_2=1.3$ $\eta=0$, $\zeta=0$, and $\tau=0$; and (b) $\chi_2=0.9, 1.1, \text{and}~ 1.7$ when $\beta=0.3$, $\eta=0$, $\zeta=0$, and $\tau=0$; (c) $\beta=0.3$, $\chi_2=0.9$, $\eta=0$, and $\xi=0$. The other plasma parameters are $\chi_1=0.025$, $\chi_3=0.8$, $\chi_4=1.3$, $\omega_{ci}=0.3$, $k=1.8$.}
\label{6Fig:F5}
\end{figure}

Now, to see the effect of physical plasma parameters on the properties of the first-order DIA rogue waves in our magneto-plasma system one has to consider the rogue wave solution of \eqref{6eq:15}. Therefore, the first-order DIA rogue wave solution of the $(3+1)$-dimensional NLSE \eqref{6eq:15}
is given by \cite{Guo2014,Abdelwahed2017,Irfan2019,Haque2019,Haque2020,Haque2021}
\begin{equation}\label{6eq:17}
\Phi=\sqrt{\frac{\lambda}{Q}}\left(\frac{\chi-8i\tau\lambda-3}{1+\chi}\right)e^{i\tau\lambda}\,,
\end{equation}
where $\chi=2\delta^2+4\tau^2\lambda^2$, $\lambda=P-2M$, $\delta=\xi+\eta+\zeta$. Figure \ref{6Fig:F5} displays how the amplitude and width of first-order DIA rogue waves in mangeto-complex plasma medium change with nonthermal parameter $\beta$ and electron to ion number density ratio $\chi_2$. The DIA rogue wave profiles against transverse perturbation $\xi$ are shown in Figures \ref{6Fig:F5}(a) and \ref{6Fig:F5}(b). The amplitude and width of rogue waves decrease with the increasing the value of $\beta$ and $\chi_2$ at higher value $k$. The spatio-temporal evolution of DIA rogue waves in $(\zeta,\tau)$-coordinates are shown in Fig. \ref{6Fig:F5}(c). It is also observed that the amplitude and width of rogue waves decrease with decreasing the values of $\chi_3$ and $\chi_4$, but the amplitude and width of rogue waves increase with reducing the value of $\chi_1$. The increasing the value of ion cyclotron frequency through the external magnetic field $B_0$ causes to increase in the energy of rogue waves in our plasma system.

\section{Summary}
\label{6sec:Discussion}
We have examined the three-dimensional modulation stability/instability of DIA waves in a magnetized dusty plasma medium containing nonthermal distributed electrons and positrons, warm adiabatic ions, and immobile dust species. We have found that the occurrence of modulation instability/stability of DIA waves in the presence of an external uniform magnetic field and obliqueness is different from the modulation instability in an unmagnetized one-dimensional case. It is observed that the product of nonlinear and dispersion coefficient $PQ$ is not sufficient to study the modulation instability in an unmagnetized one-dimensional case. To study those phenomena, we have derived a (3+1)-dimensional nonlinear Schr\"odinger equation by using the reductive perturbation method and later also a dispersion relation from that equation to find the condition for the formation of stability/instability of DIA waves. We have also examined the basic properties of DIA rogue waves in our plasma system. The main results from our present investigation are summarized as follows:
 \begin{itemize}
   \item As we increase the wave number $k$, the angular frequency of DIA waves increases with the nonthermal parameter $\beta$ but decreases with the electron-to-ion number density ratio.
   \item The group velocity of DIA waves decreases with the increase of wave propagation constant $k$. It is found that when $k$ is small, the group velocity of DIA waves will have a larger value at a higher value of the nonthermal parameter, but the opposite situation has been observed at a larger value of $k$. It means that the rate of change of group velocity with wave number is higher at the higher value of the nonthermal parameter. However, when $k$ is small, the value of $v_g$ decreases with the increases of $\chi_2$, but after a certain value of $k$ the group velocity, $v_g$ will have the larger value with increasing the value of $\chi_2$. For both cases, the group velocity reaches a constant value at a larger value $k$.
   \item We have found that the effects of plasma parameters are to change significantly the modulation instability region in magnetized plasma medium as compared to the unmagnetized case.
   \item It is seen that in the presence of an external uniform magnetic field, the basic features of first-order DIA rogue waves have been modified significantly with the physical plasma parameter. As we increase the electron temperature compared to the ion temperature the amplitude and width of the DIA rogue waves increase. On the other hand, the amplitude and width of rogue waves decrease with increasing the positron temperature in our present system.
   \item The increase in the cyclotron frequency through the external magnetic field causes to make rogue waves more energetic in our plasma system.
   \item Due to the transverse plane perturbations, the modulation properties of the DIAWs in unmagnetized and magnetized plasma systems are different.
 \end{itemize}
In conclusion, we hope that our theoretical results are capable of describing the modulation stability/instability of DIA waves and the propagation nature of the first-order DIA rogue waves in astrophysical plasma, viz., pulsar magnetospheres \cite{Michel1982}, solar wind \cite{Shrauner1979}, and magnetosphere of the Earth \cite{Gusev2000}, etc.,  and also in laboratories \cite{Shukla2002}.

\end{document}